\documentclass[%
 reprint,
superscriptaddress,
 amsmath,amssymb,
 aps,
]{revtex4-2}

\usepackage{graphicx}
\usepackage{dcolumn}
\usepackage{bm}
\usepackage[dvipsnames]{xcolor}
\usepackage{booktabs}
\usepackage{hyperref}
\usepackage{orcidlink}

\begin{document}

\preprint{APS/123-QED}

\title{The Stability of the BAO Linear Point under Modified Gravity}

\author{Jaemyoung (Jason) Lee\,\orcidlink{0000-0001-6633-9793}}
\email{astjason@sas.upenn.edu}
\affiliation{Department of Physics and Astronomy, University of Pennsylvania, Philadelphia, PA 19104, U.S.A. }
\author{Bartolomeo Fiorini\,\orcidlink{0000-0002-0092-4321}}%
\affiliation{Institute of Cosmology \& Gravitation, University of Portsmouth, Dennis Sciama Building,
Burnaby Road, Portsmouth, PO1 3FX, United Kingdom }

\author{Farnik Nikakhtar\,\orcidlink{0000-0002-3641-4366}}%
\affiliation{Department of Physics, Yale University, New Haven, CT 06511, U.S.A } 
\author{Ravi K. Sheth\,\orcidlink{0000-0002-2330-0917}}%
\affiliation{Department of Physics and Astronomy, University of Pennsylvania, Philadelphia, PA 19104, U.S.A. }%
 
\date{\today}

\begin{abstract}
Baryon Acoustic Oscillations (BAOs) are crucial in cosmological analysis, providing a standard ruler, as well as constraints on dark energy.~In General Relativity models, the BAO Linear Point -- the midpoint between the dip and the peak in the correlation function -- has been shown to be rather robust to evolution and redshift space distortions.~We show that this remains true even when the gravity model is not General Relativity, at least for $f(R)$ and DGP gravity models which have the same expansion history as the standard $\Lambda$CDM.  For the Linear Point to be able to distinguish between modified gravity (MG) and $\Lambda$CDM, survey volumes of order tens of cubic Gpc are required. 
\end{abstract}

\maketitle


\section{\label{sec:intro}Introduction}

After the astonishing discovery of the accelerating universe in the late 1990s \citep{riess1998,perlmutter1999} using Type Ia supernovae as standardizable candles, the responsible relic for this phenomenon known as dark energy has been confirmed and constrained independently by the angular distribution of photons in the Cosmic Microwave Background (CMB) \citep{Planck2018_VI} and the closely related Baryon Acoustic Oscillations (BAOs) manifest in the three-dimensional galaxy distribution at later times \citep{eisenstein2005bao}.  

The BAOs imprint a feature -- a peak and dip -- in the pair correlation function on the comoving scale $\sim 140 \text{ Mpc}$.~Most constraints on the exact size of this `standard ruler' come from fitting the predictions of a cosmological model to the measured pair counts \citep{eisenstein2005bao,eBOSS2021,abbott2024dark}. This is hampered by the fact that the BAO scale gets smoothed and shifted due to linear physics, non-linear gravitational evolution, as well as redshift-space distortions, and because we can only measure this scale using biased tracers of the underlying field.~These complications pose obstacles to sub-percent precision cosmology.~More recently, the evolution of the Hubble constant and of fluctuations, both appear to be in tension with the standard model \cite{Planck2018_VI,heymans2021kids_S8,DESY3_S8_2022,riess2022H0,brout2022pantheon+,madhavacheril2024act,DESI2024III_BAO,DESI2024VI_Cosmology}.~Therefore, it is desirable to estimate the length of the standard ruler in a way that is less tied to the standard model. 

Ref.~\citep{anselmi2016LP_BAO} argue that the midpoint between the BAO peak and dip, named the Linear Point (LP), is more robust to the effects of evolution and scale-dependent bias, at least in the standard model.  Subsequent work \citep{anselmi2018CF_standard_ruler,anselmi2018LP_mock_catalogs,anselmi2019LP_fitting,parimbelli2021LP_neutrino,nikakhtar2021BAO_Laguerre,nikakhtar2021BAO_Laguerre_mock_catalogues,anselmi2023LP_non-fiducial,he2023_BOSS_LP} has verified that the LP indeed better fits the description of a percent-level standard ruler than the dip or the peak scales, at least in the standard cosmological model and small variations from it.

One of the virtues of the LP is that it enables an estimate of the distance scale without having to fit the predictions of a specific cosmological model to the measurements.~Thus, in addition to being more robust, it potentially furnishes an estimate of the distance scale that is not as closely tied to the details of the underlying cosmological model.  Here, we study if the LP remains useful when our model of gravity is not General Relativity (GR). We do this in two steps, for two often utilized models of modified gravity:~$f(R)$ \citep{hu_sawicki2007} and nDGP \citep{DGP2000}.~Both are chosen to satisfy constraints on the expansion history, so that clustering information provides genuinely new constraints.~Whereas the first has $k$-dependent growth even in linear theory, the second is $k$-independent but has a different growth history from GR. 

First, we exhibit the correlation function shapes (CF) and the peak, dip and the linear point for these models in the limit in which the whole evolved dark matter field is observable (i.e.~with essentially no measurement noise), to see if the linear point remains a useful probe.~Then we include the degradation in the signal that comes from the fact that we only observe a sparse, biased subset of the full field, and these observations suffer from redshift-space distortions (RSD) as well as scale-dependent bias.~This allows us to check if the LP in the current generation of surveys can provide precise and accurate constraints that are not as strongly tied to GR.  

\begin{figure*}
\includegraphics[width=0.98\textwidth]{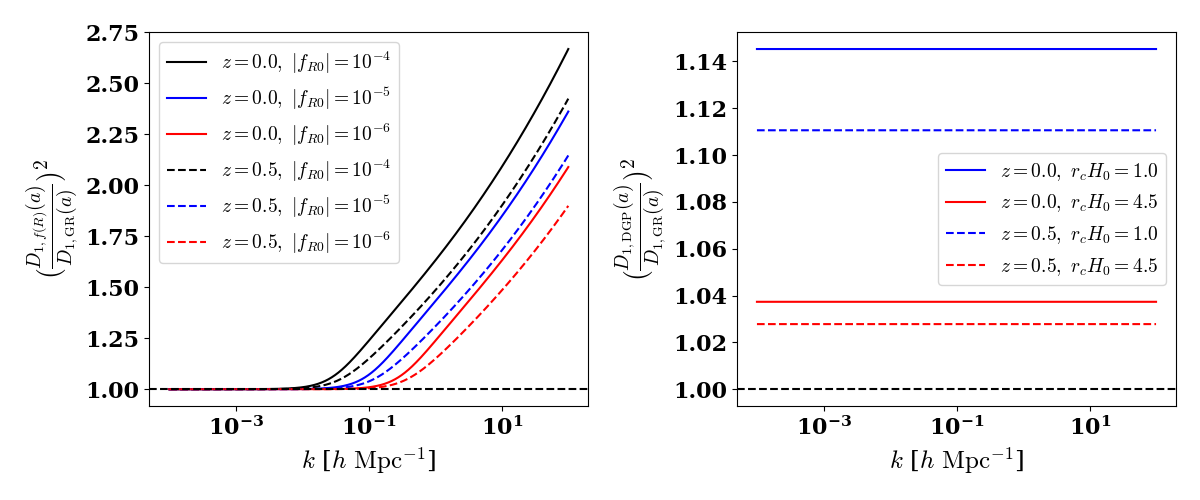}
\caption{\label{fig:growth_ratios_squared} $(\frac{D_{1, \rm{MG}}}{D_{1, \Lambda \rm{CDM}}})^2$ for $f(R)$ (left) and DGP (right).~The ratio for $f(R)$ has a scale dependence and reaches larger values, while for DGP, the ratio is constant, but not as high as the high-$k$ values for $f(R)$.~At $z = 0.5$, the ratios are smaller and $|f_{R0}| = 10^{-5}$ is similar in `strength' with $r_c H_0 = 1.0$ and $|f_{R0}| = 10^{-6}$ with $r_c H_0 = 4.5$.}
\end{figure*}

This paper is organized as follows.~In Section \ref{sec:MG_BAO}, we briefly introduce the modified gravity models and simulations we consider in this work.~We also describe the BAO two-point correlation function formalism in modified gravity. Since much of the discussion is about measurement precision, we also discuss how we estimate error bars on the BAO scales.~Next, in Section \ref{sec:results}, we show our results and compare with simulations.~Lastly, we end with a conclusion and discussion in Section \ref{sec:conclusion}.

\section{\label{sec:MG_BAO}BAO Formalism in Modified Gravity}

As noted in the Introduction (Section \ref{sec:intro}), we will explore two classes of models.~For both, we supplement measurements from the PITER simulations of COmoving Lagrangian Acceleration (COLA) in Modified Gravity \citep{fiorini2021_COLA_MG} with analytic estimates.~This is in part because we only have 5 realizations of a $\sim 1~h^{-1}$Gpc box for each model, and this does not adequately sample the cosmic variance.  

\subsection{Background expansion and linear theory growth}

The background cosmology in these PITER simulations has 
\begin{equation}
\begin{split}
    \Omega_{m,0} = 0.281,~ \Omega_{\Lambda,0} &= 0.719, ~\Omega_{b,0} = 0.046  \\
    n_s = 0.971,~ \sigma_8 &= 0.842,\quad {\rm and}\quad h = 0.697    .
\end{split}
\end{equation}

\noindent We also use the same survey volume (box size) as COLA for our default binned results, or $(1024~\mathrm{Mpc}~h^{-1})^3$, but also show forecasts for 30 times that size, which is similar to the volume we will have available with future surveys like the complete Dark Energy Spectroscopic Instrument (DESI) survey \citep{desi2016_part1}.

For our default linear theory $\Lambda$CDM $P(k)$, we use the Cosmic Linear Anisotropy Solving System (CLASS) \citep{CLASSII2011} \footnote{\url{https://lesgourg.github.io/class_public/class.html}} values at $z = 0$ and shift back to $z = 0.5057$ (also denoted as $z = 0.5$ or $z = 0.51$) and $1.0$ using the linear theory growth factors \cite[e.g.][]{hamilton2001growth}.~We assume that the MG models have the same $P(k)$ at early times (e.g. $z=100$), but differ at later times.~So, we simply multiply the $\Lambda$CDM matter power spectrum with $(D_{1, \rm{MG}}/D_{1, \Lambda \rm{CDM}})^2$ where $D_{1, \rm{MG}}$ and $D_{1, \Lambda \rm{CDM}}$ are the linear theory growth factors \cite[e.g.~Eqs.~3.5 and 4.4 of][]{winther2017cola}. Figure~\ref{fig:growth_ratios_squared} shows that, compared to $\Lambda$CDM, the $f(R)$ growth factors depend on $k$, whereas for DGP they are just a different amplitude.~This change in linear theory shape is a potential source of systematic error in analyses which assume the GR shape when estimating cosmological parameters from measurements. 

\begin{figure*}
\includegraphics[width=0.98\textwidth]{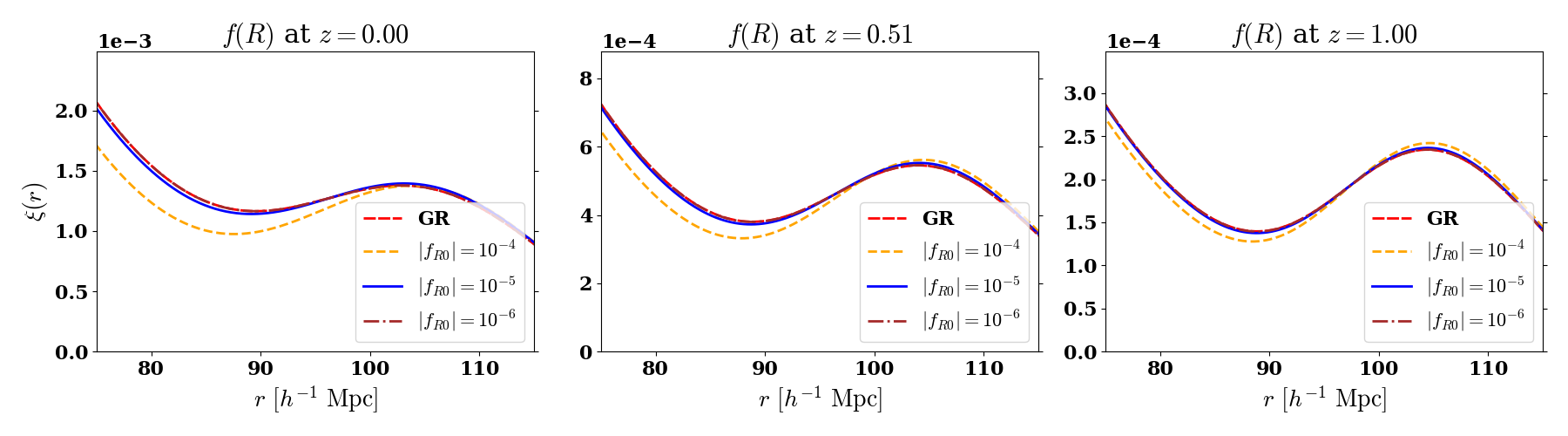}
\caption{\label{fig:xi_fR_smooth} Zeldovich smeared $\xi(r)$ curves for $f(R)$ gravity and GR. There is more deviation from GR around the dip scales compared to the peak, especially at lower redshifts.~For $z = 0.51$ and $z = 1.00$, $\xi(r_{\rm{LP}})$ values are closer to GR compared to $\xi(r_{\rm{Dip}})$ or the $\xi(r_{\rm{Peak}})$.  ~These differences depend on $f_{R0}$: while $|f_{R0}| = 10^{-6}$ curves are indistinguishable from GR, $|f_{R0}| = 10^{-4}$ curves show clear differences.}
\end{figure*}
\begin{figure*}
\includegraphics[width=0.98\textwidth]{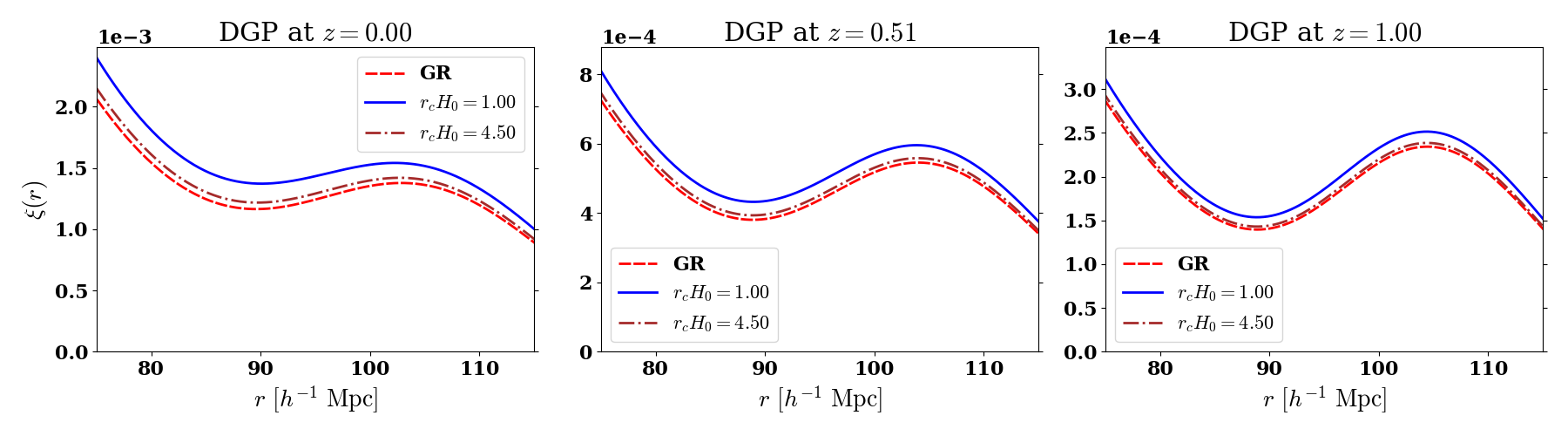}
\caption{\label{fig:xi_DGP_smooth} Same as previous figure, but for DGP. The deviation from GR is nearly just a multiplicative constant at all scales, since the growth factors are independent of scale.  In addition, differences are larger at lower redshifts and smaller $r_cH_0$.}
\end{figure*}

\subsection{Nonlinear evolution on BAO scales}
In practice, measurements are always made in the evolved field.~On BAO scales, gravitational evolution changes the shape of the pair correlation function.~The leading order correction is a smearing of the BAO feature that is caused by peculiar velocities \cite[e.g.][]{rpt2008, aviles2017LPT_MG}:   
\begin{equation}
    P^{\text{nl}}(k,z) \approx e^{-k^2\sigma_{\rm{v}}^2(z)}P^{\text{lin}}(k,z) ,
\end{equation}
where 
\begin{equation}
    \sigma_{\rm{v}}^2(z) = \frac{1}{3} \int \frac{d^3 q}{(2\pi)^3} \frac{P^{\text{lin}}(q, z)}{q^2} .
\end{equation}
Then, the `Zeldovich smeared' non-linear $\xi$ is \citep{anselmi2016LP_BAO} : 
\begin{equation}
    \xi^{\text{nl}}(r,z) 
     = \int \frac{dk}{k} \frac{k^3 P^{\text{nl}}(k,z)}{2\pi^2} \,j_0(kr) .
     \label{eq:xinl}
\end{equation}
Although the shape of $P^{\text{lin}}(k)$ in DGP and $\Lambda$CDM is the same, the difference in growth factors means that the smearing will be slightly different, so $\xi^{\rm nl}$ can differ.  Figures~\ref{fig:xi_fR_smooth} and~\ref{fig:xi_DGP_smooth} show the magnitude of the effect; our results for $f(R)$ are similar to those shown in Figure~8 of \citep{aviles2017LPT_MG}.~(In principle, there are `mode-coupling' corrections to this approximation; these matter more on smaller scales than the ones of interest here.)

\subsection{Linear point in idealized conditions}
The agreement with previous work is reassuring, as our main goal here is to quantify the stability of the linear point scale $r_{\rm LP}$ in these MG models.~Following \cite{anselmi2016LP_BAO}, to find the LP, we find where $d\xi/dr = 0$ to obtain $r_{\text{peak}}$ and $r_{\text{dip}}$, and then set  
\begin{equation}
    r_{\text{LP}} \equiv \frac{r_{\text{peak}} + r_{\text{dip}}}{2}.
\end{equation}\label{eq:LP_def}
\noindent Clearly, $r_{\rm LP}$ will depend on the value of $\sigma_{\rm{v}}$: the `standard rod' scale is when $\sigma_{\rm{v}}=0$, so we will denote it as $r_{\rm LP0}$. In contrast, the LP in $\xi^{\text{nl}}$ will be slightly different for non-zero $\sigma_{\rm{v}}$ (especially if $\sigma_{\rm{v}}$ is large).~\citep{anselmi2016LP_BAO} note that a crude way to mitigate this effect of evolution is to multiply $r_{\text{LP}}$ by 1.005 (a 0.5\% correction).~This was motivated by the evolution seen in simulations of $\Lambda$CDM, so it is not obviously appropriate for modified gravity models.~\cite{nikakhtar2021BAO_Laguerre,nikakhtar2021BAO_Laguerre_mock_catalogues} describe a slightly more elaborate way to reconstruct the linear LP from measurements of the evolved one.~We will comment on both approaches later.

\begin{table}[h!]
    \centering
\begin{tabular}{cccc}
\toprule
Type & $r_{\mathrm{LP}}/r_{\mathrm{LP0}}$  & $r_{\mathrm{Dip}}/r_{\mathrm{Dip0}}$  & $r_{\mathrm{Peak}}/r_{\mathrm{Peak0}}$  \\
\midrule
GR & 0.991 & 0.999 & 0.984 \\
$|f_{R0}| = 10^{-4}$ & 0.985 & 0.976 & 0.993 \\
$|f_{R0}| = 10^{-5}$ & 0.989 & 0.993 & 0.986 \\
$|f_{R0}| = 10^{-6}$ & 0.991 & 0.999 & 0.984 \\
$r_cH_0 = 1.0$ & 0.990 & 1.004 & 0.978 \\
$r_cH_0 = 4.5$ & 0.990 & 1.000 & 0.982 \\
\bottomrule
\end{tabular}
    \caption{BAO scales at $z = 0.0$ (as fractions of linear theory values) for different variations of MG. The LT BAO scales are: $[r_{\rm{LP0}},~ r_{\rm{Dip0}},~r_{\rm{Peak0}}] = [97.154,~89.699,~104.609]~ h^{-1}\rm{Mpc}$. As expected, $r_{\rm{LP}}$ is the most stable, showing smaller deviations from the GR values compared to $r_{\rm{Dip}}$ and $r_{\rm{Peak}}$. For $|f_{R0}| = 10^{-4}$ and $10^{-5}$, $r_{\rm{Dip}}$ shifts to the left and $r_{\rm{Peak}}$ shifts to the right, with the net effect being a small shift to the left for $r_{\rm{LP}}$.~For DGP, the opposite is true; $r_{\rm{Dip}}$ shifts to the right, while $r_{\rm{Peak}}$ shifts to the left, although by more subtle amounts compared to $f(R)$.}
    \label{tab:r_scales_smooth_z0}
\end{table}

\begin{table}[h!]
    \centering
\begin{tabular}{cccc}
\toprule
Type & $r_{\mathrm{LP}}/r_{\mathrm{LP0}}$ & $r_{\mathrm{Dip}}/r_{\mathrm{Dip0}}$ & $r_{\mathrm{Peak}}/r_{\mathrm{Peak0}}$ \\
\midrule
GR & 0.993 & 0.991 & 0.994 \\
$|f_{R0}| = 10^{-4}$ & 0.990 & 0.980 & 0.999 \\
$|f_{R0}| = 10^{-5}$ & 0.993 & 0.989 & 0.995 \\
$|f_{R0}| = 10^{-6}$ & 0.993 & 0.991 & 0.994 \\
$r_cH_0 = 1.0$ & 0.993 & 0.992 & 0.993 \\
$r_cH_0 = 4.5$ & 0.993 & 0.991 & 0.994 \\
\bottomrule
\end{tabular}
    \caption{Same as Table \ref{tab:r_scales_smooth_z0}, but for $z = 0.5057$. Now, the MG and GR BAO scales are either the same or nearly so with the exception of $|f_{\rm{R0}}| = 10^{-4}$ and $10^{-5}$. The shifts for $f(R)$ display similar trends as at $z = 0.0$, with $r_{\mathrm{Dip}}$ shifting left and $r_{\mathrm{Peak}}$ shifting right.}
    \label{tab:r_scales_smooth_zhalf}
\end{table}

\begin{table}[h!]
    \centering
\begin{tabular}{cccc}
\toprule
Type & $r_{\mathrm{LP}}/r_{\mathrm{LP0}}$ & $r_{\mathrm{Dip}}/r_{\mathrm{Dip0}}$ & $r_{\mathrm{Peak}}/r_{\mathrm{Peak0}}$ \\
\midrule
GR & 0.994 & 0.991 & 0.998 \\
$|f_{R0}| = 10^{-4}$ & 0.993 & 0.985 & 1.000 \\
$|f_{R0}| = 10^{-5}$ & 0.995 & 0.991 & 0.999 \\
$|f_{R0}| = 10^{-6}$ & 0.994 & 0.991 & 0.998 \\
$r_cH_0 = 1.0$ & 0.994 & 0.991 & 0.998 \\
$r_cH_0 = 4.5$ & 0.994 & 0.991 & 0.998 \\
\bottomrule
\end{tabular}
    \caption{Same as Table \ref{tab:r_scales_smooth_z0}, but for $z = 1.0$. Here, we only see some difference between the BAO scales for $|f_{\rm{R0}}| = 10^{-4}$. The shift directions for all three scales are the same as at $z = 0.0$ and $z = 0.5057$.}
    \label{tab:r_scales_smooth_z1}
\end{table}

In Tables \ref{tab:r_scales_smooth_z0}-\ref{tab:r_scales_smooth_z1}, we show the BAO scales for $z = 0.0$, $z = 0.5057$, and $z = 1.0$ for the different variations of MG we consider in this work as fractions of linear theory (LT) BAO scales which are: $[r_{\rm{LP0}},~ r_{\rm{Dip0}},~r_{\rm{Peak0}}] = [97.154,~89.699,~104.609]~h^{-1}\rm{Mpc}$.  
As with Figs.~\ref{fig:xi_fR_smooth} and~\ref{fig:xi_DGP_smooth}, we see the most deviation from GR values at $z = 0.0$. Using Table~\ref{tab:r_scales_smooth_z0} as reference since the shifts are largest, $r_{\rm{LP}}$ is shown to be the most stable for each MG variation, with the exception of $|f_{R0}| = 10^{-6}$ where the shifts are extremely small anyway.~Compared to GR, the $f(R)$ models have smaller $r_{\rm{Dip}}$ and larger $r_{\rm{Peak}}$, so that $r_{\rm LP}$ ends up slightly smaller.~This is opposite to the DGP models, which have slightly larger $r_{\rm{Dip}}$ and smaller $r_{\rm{Peak}}$ than GR.  

Tables \ref{tab:r_scales_smooth_zhalf} and \ref{tab:r_scales_smooth_z1} show that at higher redshifts where the fluctuations were smaller, the MG BAO scales become more similar to GR values.~Based on Tables \ref{tab:r_scales_smooth_z0}-\ref{tab:r_scales_smooth_z1}, it is evident that $r_{\rm{LP}}$ is more stable compared to either $r_{\rm{Dip}}$ or $r_{\rm{Peak}}$ under the modified gravity models we explore.~However, the shifts are typically less than 0.5\% of the LT values, suggesting that determining the correct MG model given current observational constraints will be challenging. 

This conclusion is slightly pessimistic, because measurements will suffer from redshift space distortions which increase the overall amplitude of $\xi$ \cite[e.g.][]{kaiser1987rsd}, potentially increasing the signal to noise of a measurement, and further smear the BAO feature (see below), potentially leading to additional shifting of $r_{\rm Peak}$, $r_{\rm Dip}$ and $r_{\rm LP}$.  We discuss this in more detail below.  

\subsection{\label{sec:Bias_factor}Biased tracers and redshift space distortions}

The galaxies we observe are biased and redshift-space distorted tracers.~We account for this by modelling the reshift-space distorted monopole and quadrupole of the biased tracers as 
\begin{equation}
    \xi_\ell^{\rm nl}(s) = \frac{i^\ell (2\ell + 1)}{2} \int_{-1}^{1} \mathcal{L}_\ell(\mu) d\mu \int \frac{dk}{k} \Delta_b^{\text{nl}}(k,\mu)\,j_\ell (ks)
\label{eq:xil_def}
\end{equation}
where
\begin{equation}
    \Delta_b^{\text{nl}}(k,\mu) = \frac{k^3P^{\text{nl}}(k)}{2\pi^2}\, (b + \mu ^2 f)^2\,
                          e^{- k^2 \sigma_{\rm{v}}^2 \mu ^2 f (2 + f)}
\end{equation}
\cite[e.g.][]{paranjape2023}.~Here, $\mathcal{L}_\ell(\mu)$ is the $\ell$th Legendre polynomial, $j_\ell(x)$ is a spherical Bessel function, $f \equiv d\ln D/d\ln a$, and $b$ is the `linear bias factor' (Eq.~\ref{eq:xinl} is this expression with $\ell=0$, $f=0$ and $b=1$).  

This raises the question of what to use for $b$.~If we restrict attention to all halos above some minimum mass (set, e.g., by a cut on mass or number density), then their clustering strength -- hence the value of $b$ -- may well be different for each background model.~In principle, this can arise because both $b_{\rm MG}/b_{\rm GR}$ and $D_{\rm MG}/D_{\rm GR}$ might depend on scale, and allowing for both $k$-dependent bias as well as $k$-dependent growth might result in an enhanced MG signal.~In practice, however, when we observe a set of galaxies, there is an unknown transformation from `halo' to `galaxy' statistics that is set by gastrophysics.~Since the gastrophysics is relatively unconstrained (compared to the precision with which cosmological parameters are known), we have taken a more conservative approach which we believe will yield a more realistic estimate of the actual constraining power of biased data sets.~We assume that, no matter what the background model, it must give rise to the observed number density and real-space two-point clustering signal $w_p(r_p)$. We approximate this constraint by requiring that the real space clustering strength $b^2(k)\,P(k)$ is the same as in GR, with modifications from MG only arising from differences in $f$ and $\sigma_{\rm v}$.  

In the following section, we perform a more in-depth analysis which includes this effect, as well as the impact of having a finite survey volume, so that the pair correlation function must be estimated by counting pairs in bins of non-negligible width.  

\begin{figure*}
\includegraphics[width=0.98\textwidth]{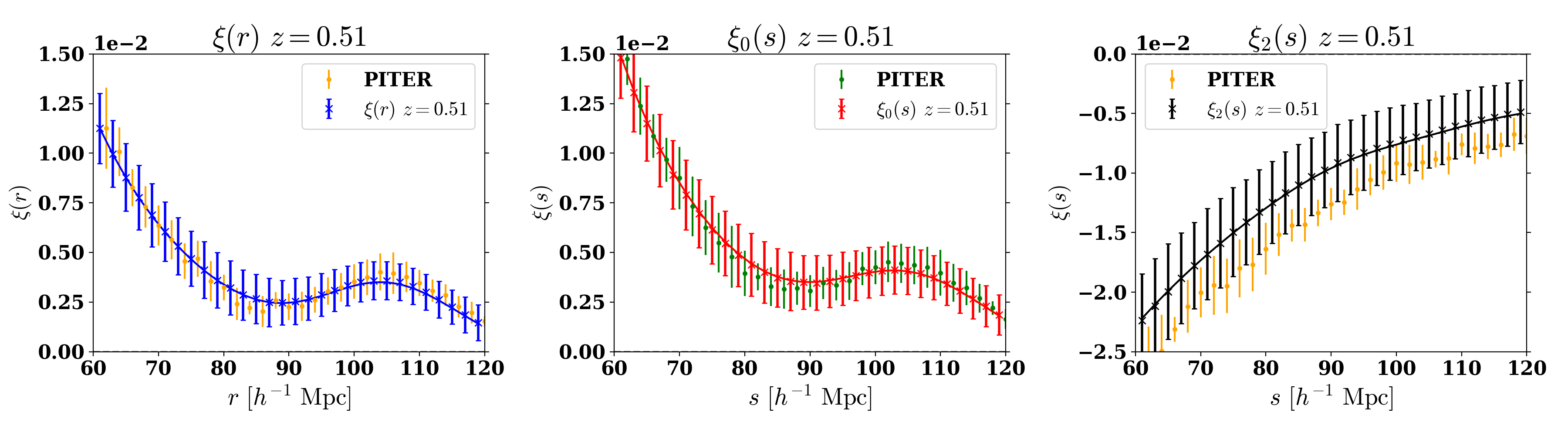}
\caption{\label{fig:xi_GR_100} Measurements of $\xi(r)$ (left), $\xi_0(s)$ (middle) and $\xi_2(s)$ (right) in our GR model: error bars show the RMS of 5 PITER and 100 mock realizations (effective volume $\sim 1~h^{-3}\rm{Gpc}^3$), respectively.~PITER values have been displaced by $1.0~h^{-1}$ Mpc to the right for clarity.}
\end{figure*}

\subsection{\label{sec:BAO_Covariance}Biased tracers in a finite survey volume}
As noted above, measurements of the TPCF in finite observed datasets are made by binning the pair counts into bins of non-vanishing width.  The counts in different bins are not independent:  their correlation is quantified by a covariance matrix $C^\xi_{\ell_1\ell_2}(s_i,s_j)$ which describes how different multipole counts in a bin differ from each bin's mean value, where the mean is the expected signal if the survey volume were infinite.~For a total survey volume $V_s$ containing biased tracers with a number density $\bar{n}_b$ and expected clustering strength $\Delta_b^{\rm nl}(k,\mu)$, this covariance is well described by the `Gauss-Poisson' approximation \citep{sss2008bao_motion, grieb2016gaussian_covariance, parimbelli2021LP_neutrino}, at least on the large scales of relevance to BAO studies.  Namely, 
\begin{equation}
    C_{\ell_1 \ell_2}^{\xi} (s_i,s_j) = \frac{i^{\ell_1 + \ell_2}}{2\pi^2} \int_{0}^{\infty} k^2 \sigma_{\ell_1\ell_2}^2(k) \bar{j}_{\ell_1}(ks_i)\bar{j}_{\ell_1}(ks_j)dk
\label{eq:TPCF_cov_mat}
\end{equation} 
\noindent where $V_s$ is the survey volume, $\ell_1$ and $\ell_2$ are the multipoles, $\bar{n}$ is the shot noise, and $\mathcal{L}$ are Legendre polynomials, and 
\begin{equation}
    \begin{split}
        \sigma_{\ell_1\ell_2}^2(k) & \equiv \frac{(2\ell_1+1)(2\ell_2+1)}{V_s} \\
        & \times \int_{-1}^{1} \big[ P(k, \mu) + \frac{1}{\bar{n}}\big]^2\mathcal{L}_{\ell_1}(\mu)\mathcal{L}_{\ell_2}(\mu)  d\mu
    \end{split}
\end{equation}
is the multipole expansion of the per-mode covariance.  

\begin{figure*}
\includegraphics[width=0.95\textwidth]{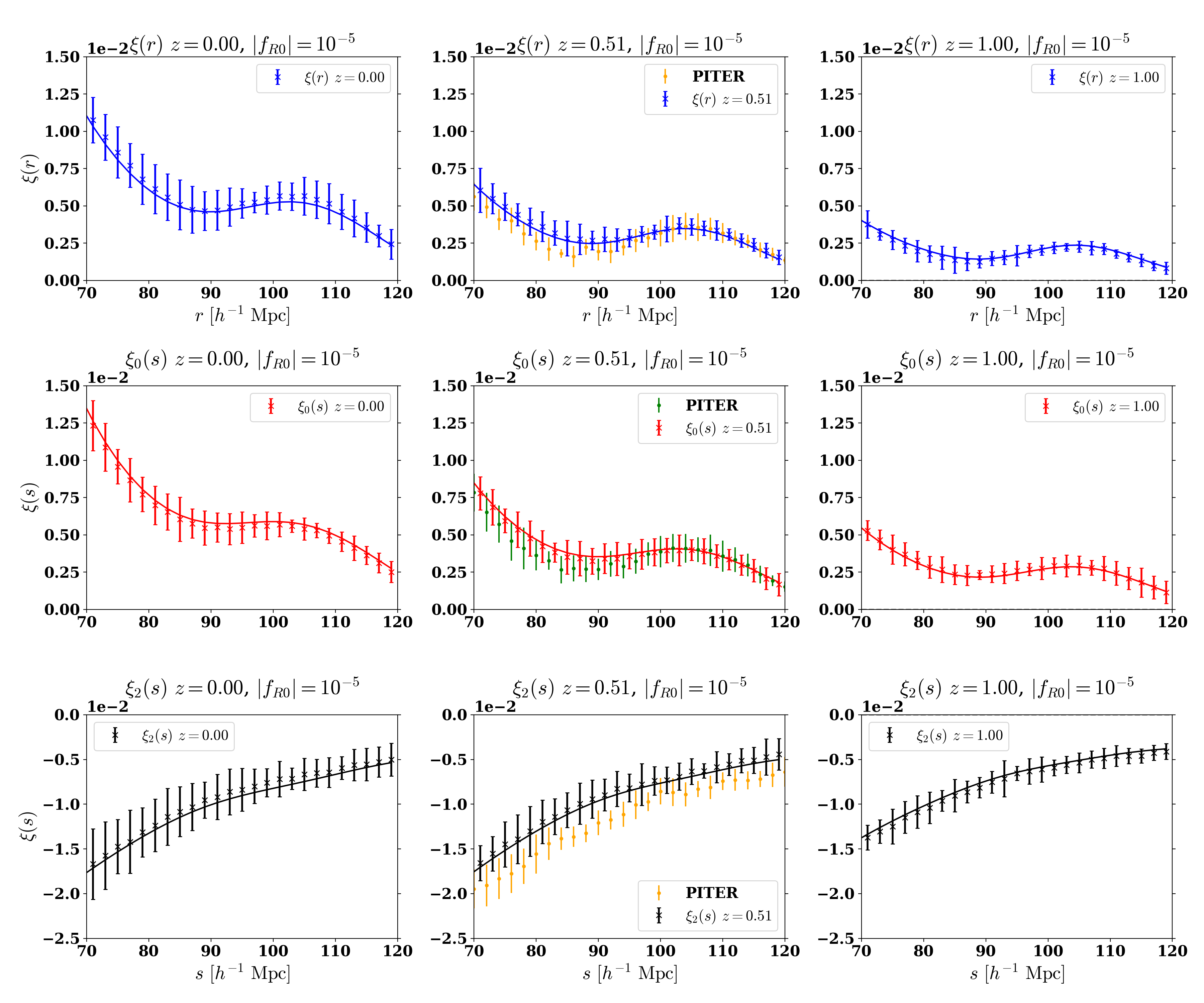}
\caption{\label{fig:xi_fR_PITER} Analytic and PITER binned $\xi(r)$ (top panels), $\xi_0(s)$ (middle panels), and $\xi_2(s)$ (bottom panels) with $1\sigma$ error bars for $f(R)$ gravity with $|f_{R0}| = 10^{-5}$. The solid lines are theoretical curves while the symbols and error bars are the binned values obtained from the analytic covariance matrices. The PITER values have been displaced by $1.0~h^{-1}$ Mpc to the right for clarity. We see good agreement between the theoretical values and the binned values.~The binned values and PITER values agree reasonably for $\xi(r)$ and $\xi_0(s)$ with the $p$-value from the $\chi^2$ being above 0.10 but more so for $\xi_0(s)$ and $\xi(r)$ at $r ~(\mathrm{or}~s) > 95~h^{-1}$ Mpc. We see some discrepancy for $\xi_2(s)$ with the $p$-value from the $\chi^2$ being below 0.01, which could be due to mode coupling terms we have not incorporated into our analysis. The trends seen here are similar for GR and DGP gravity.}
\end{figure*}

We use this to generate a mock realization of the TPCF made in such a survey as follows.~First we diagonalize $C^\xi(s_i,s_j)$.~If this was an $N_{\rm bin}\times N_{\rm bin}$ matrix, then its eigenvectors, which we denote $\Lambda_i(s)$, provide an orthogonal set of $N_{\rm bin}$ shape functions. Therefore, we can write one realization $\Xi$ of the TPCF as: 
\begin{equation}
    \Xi(s) = \xi(s) + \sum_{i=1}^{N_{\rm bin}} g_i \Lambda_i(s), 
\label{eq:Xi_realization}
\end{equation}
where $\xi(s)$ is given by equation~(\ref{eq:xil_def}) and the $g_i$'s are independent Gaussian random variates with zero mean and variance equal to the corresponding eigenvalue $\lambda_i$. Hence, our second step is to generate $N_{\rm bin}$ independent $g_i$ and insert in equation~(\ref{eq:Xi_realization}).  (We do this for each $\ell$, and, if necessary, we could have accounted for the covariance between different $\ell$s.)  By repeating this procedure many times we can generate many mock realizations of each $\xi_\ell$, from which we can estimate the mean, and for which the scatter between realizations is given by the Gauss-Poisson model.  

In addition, for each realization, we can estimate $r_{\mathrm{Peak}}$ and $r_{\mathrm{Dip}}$ and hence $r_{\rm LP}$, and from the values in different realizations, we can estimate error bars on $r_{\rm LP}$.  In practice, we estimate $r_{\mathrm{Peak}}$ and $r_{\mathrm{Dip}}$ in each realization by fitting the TPCF, estimated in bins that are $2~h^{-1}$Mpc wide, over the range $[60,120]~h^{-1}$Mpc to a 7th order polynomial \cite[following, e.g.][]{anselmi2018LP_mock_catalogs, paranjape2023}. 

\section{\label{sec:results}Results}

To illustrate our approach, we assume the number density and clustering of tracers is that of halos more massive than about $10^{13}\ h^{-1}M_\odot$ at $z=0.5057$ in our fiducial $\Lambda$CDM model:  $\bar{n} = 3.2 \times 10^{-4}\ h^3$Mpc$^{-3}$.  In the PITER simulations, they have a clustering strength that is approximately $b^2\, P(k)$ with $b=1.97$ \cite{fiorini2021_COLA_MG}.  
We first assume a survey volume of just $1024^3~h^{-3}$Mpc$^3$, since this is the volume of a PITER simulation box. Later, we will consider larger volumes.~We assume the same $\bar{n}$ and $b$ at $z=0$ and $1$, because our main goal -- to illustrate the approximate level of precision we can expect in upcoming datasets -- does not depend strongly on these choices. 

Figure~\ref{fig:xi_GR_100} compares the mean and root-mean-square (RMS) values of $\xi(r)$, $\xi_0(s)$ and $\xi_2(s)$ in 100 realizations of our mocks with those in the PITER simulations.~For this comparison, we have chosen the GR model at $z=0.5$.~At least for $\xi(r)$ and $\xi_0$, the means are in reasonably good agreement with one another, and with the expected shapes (equation~\ref{eq:xil_def}, with $f$ and $\ell=0$, just $\ell=0$ and $\ell=2$).~However, the PITER error bars are obviously smaller:~Evidently, 5 realizations are not enough to adequately sample the cosmic variance of $1024^3~h^{-3}$Mpc$^3$ volumes.~Some of the smaller systematic differences in the mean values arise from our neglect of mode-coupling -- which should be quite small -- and scale dependent bias, which may not be so small \cite{bkPeaks2010}. The quadrupole $\xi_2$ is off by a larger amount \cite[consistent with some previous work, e.g.][]{paranjape2023}, so, although we also show it for the MG models, we will not use it further.  

\begin{table}
\begin{tabular}{cccc}
\toprule
Type & $r_{\mathrm{LP}}/r_{\mathrm{LP0}}$ & $r_{\mathrm{Dip}}/r_{\mathrm{Dip0}}$ & $r_{\mathrm{Peak}}/r_{\mathrm{Peak0}}$ \\
\midrule
GR & 0.9922$\pm$0.0073 & 1.0136$\pm$0.0182 & 0.9738$\pm$0.0167 \\
$|f_{R0}| = 10^{-4}$ & 0.9866$\pm$0.0087 & 1.0187$\pm$0.0254 & 0.9592$\pm$0.0255 \\
$|f_{R0}| = 10^{-5}$ & 0.9951$\pm$0.0079 & 1.0203$\pm$0.0183 & 0.9735$\pm$0.0164 \\
$|f_{R0}| = 10^{-6}$ & 0.9909$\pm$0.0069 & 1.0081$\pm$0.0146 & 0.9761$\pm$0.0132 \\
$r_cH_0 = 1.0$ & 0.9933$\pm$0.0069 & 1.0173$\pm$0.0157 & 0.9728$\pm$0.0138 \\
$r_cH_0 = 4.5$ & 0.9997$\pm$0.0085 & 1.0212$\pm$0.0197 & 0.9814$\pm$0.0165 \\
\bottomrule
\end{tabular}
    \caption{BAO scales for binned TPCF at $z = 0.0$ (as fractions of linear theory values, effective volume of about $5~h^{-3}\rm{Gpc}^3$), for $\xi_0(s)$ in GR and the MG variations considered in this work.~Recall that  $[r_{\rm{LP0}},~ r_{\rm{Dip0}},~r_{\rm{Peak0}}] = [97.154,~89.699,~104.609]~ h^{-1}\rm{Mpc}$. The error bars are larger than the shifts we see in the central values of the BAO scales, suggesting that distinguishing MG models through measurement of the BAO scales is difficult using current observations. However, $r_{\mathrm{LP}}$ error bars are smaller than for $r_{\mathrm{Dip}}$ or $r_{\mathrm{Peak}}$, which implies that $r_{\mathrm{LP}}$ can be measured more precisely. 
    }
    \label{tab:r_scales_binned_z0}
\end{table}

Fig.~\ref{fig:xi_fR_PITER} shows $\xi(r)$ (top panels), $\xi_0(s)$ (middle panels), and $\xi_2(s)$ (bottom panels) for $f(R)$ gravity with $|f_{R0}| = 10^{-5}$, for the same three redshifts as in the previous section.~The trends we see for $f(R)$ are similar for GR and DGP, so we so not show them all here.~The smooth curves show equation~(\ref{eq:xil_def}), with $f$ and $\ell=0$ (top), just $\ell=0$ (middle) and $\ell=2$ (bottom), and the mean of the binned values are generally consistent with each other.~Here, the error bars show the RMS of 5 realizations, both for our mocks and for PITER.~(Recall from the previous figure that the true cosmic variance induced error bar would be larger.)

Tables~\ref{tab:r_scales_binned_z0} and~\ref{tab:r_scales_binned_zhalf} show the fractional error on the distance scale from these biased tracers if we average together the results from 5 realizations, each of volume $1024~h^{-3}$Mpc$^3$.~First, note that the error bars for the peak and dip scales are always larger than for the LP.~Nevertheless, the fractional errors on the LP scale are substantial -- $\pm 0.8$ percent -- comparable to the offset in the mean value.


The middle panels also show corresponding measurements in the PITER simulations.~Some of the disagreement between the PITER and the mock-based results is due to cosmic variance: we only have 5 PITER realizations, and the cosmic variance on BAO scales from a total volume of just $5~h^{-3}$Gpc$^3$ is still substantial (c.f. Figure~\ref{fig:xi_GR_100}). In addition, the minimum halo mass for PITER is $\sim 10^{13} M_{\odot}$.~On the other hand, our mocks ignore both mode-coupling -- which should be small on these scales -- and scale-dependent bias, which may not be negligible for this range of masses.~However, some of the disagreement arises from our choice to force the MG mocks to have the same $b^2P(k)$ as in GR -- the PITER shape is noticeably different.~E.g., Table~\ref{tab:r_scales_binned_zhalf_PITER} gives the BAO scales for the PITER simulations at $z = 0.5057$, estimated using the same covariance matrices as our mocks.~While the errors are smaller, perhaps because of the missing cosmic variance, the mean values of $r_{\rm Dip}$ and hence the $r_{\rm LP}$ have shifted to significantly smaller scales than in Table~\ref{tab:r_scales_binned_zhalf}.~Another potential explanation for this shift is our choice of not including scale-dependent bias or non-linear halo bias for our mocks.~Scale-dependent bias could affect the $r_{\rm Dip}$ scales similar to the way scale-dependent growth does to the smooth $f(R)$ curves shown in Figure~\ref{fig:xi_fR_smooth} and Tables~\ref{tab:r_scales_smooth_z0}-\ref{tab:r_scales_smooth_z1}, where $r_{\rm Dip}$ shifts to considerably smaller scales from GR.~Although this suggests that these shifts are significant compared to the measurement errors, until the halo-galaxy connection is better understood, the values in Table~\ref{tab:r_scales_binned_zhalf}, based on our synthetic mocks, are probably more realistic.~These indicate that, although the LP is both more accurate and more precise, a measurement of it in an effective volume of $5~h^{-3}$Gpc$^3$ volume is unable to distinguish between different MG models.  

\begin{table}
\begin{tabular}{cccc}
\toprule
Type & $r_{\mathrm{LP}}/r_{\mathrm{LP0}}$ & $r_{\mathrm{Dip}}/r_{\mathrm{Dip0}}$ & $r_{\mathrm{Peak}}/r_{\mathrm{Peak0}}$ \\
\midrule
GR & 0.9860$\pm$0.0080 & 0.9896$\pm$0.0141 & 0.9829$\pm$0.0121 \\
$|f_{R0}| = 10^{-4}$ & 0.9898$\pm$0.0083 & 0.9956$\pm$0.0140 & 0.9849$\pm$0.0134 \\
$|f_{R0}| = 10^{-5}$ & 0.9936$\pm$0.0076 & 0.9992$\pm$0.0133 & 0.9888$\pm$0.0112 \\
$|f_{R0}| = 10^{-6}$ & 0.9954$\pm$0.0079 & 1.0017$\pm$0.0139 & 0.9899$\pm$0.0114 \\
$r_cH_0 = 1.0$ & 0.9916$\pm$0.0077 & 1.0014$\pm$0.0139 & 0.9832$\pm$0.0115 \\
$r_cH_0 = 4.5$ & 0.9948$\pm$0.0088 & 1.0027$\pm$0.0163 & 0.9880$\pm$0.0129 \\
\bottomrule
\end{tabular}
    \caption{Same as Table~\ref{tab:r_scales_binned_z0}, but for $z = 0.5057$.\\ }
    \label{tab:r_scales_binned_zhalf}
\end{table}

Future surveys, like DESI, will observe a much larger volume. To model such a survey, we have repeated our analysis after setting $V_s$ to be about $32~h^{-3}$Gpc$^3$ or 30 times the volume of the PITER box.~This makes no difference to the estimated LP scale, but decreases the fractional uncertainty on the estimate to about 0.3 percent.~As a result, the shifts of the BAO scales in the various MG scenarios compared to the GR values are comparable to the size of the error bars (except for the case of real-space at $z = 1.0$).~For example, for $\xi_0(s)$ in DGP with $r_cH_0 = 1.0$ the central value of $r_{\rm{Dip}}$ shifts by 1.0\% to larger scales, while $r_{\rm{Peak}}$ shifts 0.4\% to smaller scales, shifting the central value of $r_{\rm{LP}}$ by only 0.2\%, as shown in Table \ref{tab:r_scales_binned_zhalf_30Gpc}. I.e., the $r_{\rm{Dip}}$ and $r_{\rm{Peak}}$ central value shifts are more comparable to the size of the error bars in this case, whereas $r_{\rm LP}$ is stable.~Note that, only when the errors are this small will systematic differences between the distance scales returned by simply applying a secular 1.005 multiplicative shift to the measured $r_{\rm LP}$ \cite{anselmi2016LP_BAO} and the slightly more elaborate Laguerre reconstruction approach of \cite{nikakhtar2021BAO_Laguerre_mock_catalogues} matter. 

\section{\label{sec:conclusion}Discussion and Conclusion}
We studied whether the BAO linear point (LP), a more stable standard ruler than the BAO Dip or the Peak under non-linearities in GR models, is also more stable in modified gravity models. 
For MG models that are constrained to have the same expansion history as GR -- only the growth of gravitational instabilities is modified -- we found that the LP is indeed more stable in MG, at least for the ideal case of unbiased tracers in an essentially infinite volume survey.~For the more realistic case of rare, biased tracers in a finite survey, the uncertainties on BAO scale estimates for current-generation surveys are too large to be able to distinguish between MG models and GR.~However, a future survey with a few tens of times the volume could reach the precision where the shifts of the BAO scales with respect to GR under MG are statistically significant.~For this purpose, the LP should be a useful workhorse, as it can be measured more precisely than the Dip or the Peak scales (Tables~\ref{tab:r_scales_binned_z0}-~\ref{tab:r_scales_binned_zhalf_30Gpc}).  

\begin{table}
\begin{tabular}{cccc}
\toprule
Type & $r_{\mathrm{LP}}$ & $r_{\mathrm{Dip}}$ & $r_{\mathrm{Peak}}$ \\
\midrule
GR & 0.9761$\pm$0.0054 & 0.9671$\pm$0.0089 & 0.9839$\pm$0.0079 \\
$|f_{R0}| = 10^{-5}$ & 0.9777$\pm$0.0052 & 0.9653$\pm$0.0085 & 0.9883$\pm$0.0072 \\
$r_cH_0 = 1.0$ & 0.9765$\pm$0.0053 & 0.9702$\pm$0.0087 & 0.9820$\pm$0.0078 \\
\bottomrule
\end{tabular}
\caption{Same as Table~\ref{tab:r_scales_binned_zhalf} but for measurements from the average of the 5 PITER realizations (effective volume of about $5~h^{-3}\rm{Gpc}^3$) at $z = 0.5057$. 
 }
\label{tab:r_scales_binned_zhalf_PITER}
\end{table}

\begin{table}
\begin{tabular}{cccc}
\toprule
Type & $r_{\mathrm{LP}}/r_{\mathrm{LP0}}$ & $r_{\mathrm{Dip}}/r_{\mathrm{Dip0}}$ & $r_{\mathrm{Peak}}/r_{\mathrm{Peak0}}$  \\
\midrule
GR & 0.9897$\pm$0.0034 & 0.9962$\pm$0.0061 & 0.9842$\pm$0.0048 \\
$|f_{R0}| = 10^{-4}$ & 0.9919$\pm$0.0034 & 1.0019$\pm$0.0060 & 0.9833$\pm$0.0048 \\
$|f_{R0}| = 10^{-5}$ & 0.9937$\pm$0.0037 & 1.0042$\pm$0.0067 & 0.9846$\pm$0.0052 \\
$|f_{R0}| = 10^{-6}$ & 0.9941$\pm$0.0034 & 1.0034$\pm$0.0062 & 0.9861$\pm$0.0049 \\
$r_cH_0 = 1.0$ & 0.9921$\pm$0.0036 & 1.0059$\pm$0.0066 & 0.9802$\pm$0.0054 \\
$r_cH_0 = 4.5$ & 0.9904$\pm$0.0032 & 0.9981$\pm$0.0060 & 0.9838$\pm$0.0047 \\
\bottomrule
\end{tabular}
    \caption{Same as Table~\ref{tab:r_scales_binned_zhalf}, but for a roughly $32~h^{-3}\mathrm{Gpc}^3$ survey.\\ }
    \label{tab:r_scales_binned_zhalf_30Gpc}
\end{table}

We argued that, unless the bias between the observed tracers and the underlying dark matter field is extremely well understood, it is reasonable to require that GR and MG models be normalized to produce some observed clustering signal,~which we chose to be a signal that is not redshift-space distorted, the real space clustering strength.~This significantly reduces the potential differences between GR and MG signals in redshift-space distorted datasets (see Tables~\ref{tab:r_scales_binned_zhalf} and~\ref{tab:r_scales_binned_zhalf_PITER}).  



\begin{acknowledgments}
We sincerely thank Carola Zanoletti for pointing out an error in the DGP growth factors in the initial version of this paper. 

JL was supported by DOE grant DE-FOA-0002424 and NSF grant AST-2108094.~FN gratefully acknowledges support from the Yale Center for Astronomy and Astrophysics Prize Postdoctoral Fellowship.~BF is supported by a Royal Society Enhancement Award (grant no. RF$\backslash$ERE$\backslash$210304).
\end{acknowledgments}

\bibliography{apssamp}

\end{document}